\documentstyle[preprint,tighten,aps]{revtex}
\begin{document}


\draft
\title{Post-Inflationary Reheating}
\author{A. B. Henriques}
\address{Departamento de Fisica/CENTRA,Instituto Superior Tecnico,
1096 Lisbon, Portugal}
\author{R. G. Moorhouse}
\address{Department of Physics and Astronomy,University of Glasgow,
Glasgow G12 8QQ, U.K.}
\date{\today}
\maketitle
\begin{abstract}
 We study the two field model for reheating based on the scalar inflaton 
 field, $\varphi$, and its interaction with another scalar field, $\chi$,  
 through a Lagrangian term $\frac{1}{2}g^2 \chi^2 \varphi^2$,much 
 investigated   for parametric resonance or preheating.
 Attention is particularly on the quantum excitations of the inflaton 
 field and the metric with a smooth transition from quantum to classical 
 stochastic states, and with reheating followed right through 
 from a specific inflation model to a state including a  
 relativistic (radiation) fluid. The excitations of the metric (but not
 the inflaton or $\chi$) are treated perturbatively and the validity  
 of this approach is assessed. For this model  our work points to the 
 possibility of $\zeta$ (the curvature associated parameter for observed 
 cosmic microwave background radiation anisotropies) 
 changing significantly during reheating with parametric resonance. 

\end{abstract} 
\narrowtext
\section{INTRODUCTION}\label{INTRO}

Cosmic inflation theory combined with the theory of quantum fluctuations
of the inflaton field has provided a theoretical basis for the 
observed cosmic microwave background fluctuations (CMBRF) \cite{LL}. 
 In its simplest form with just one scalar field $\varphi$, the inflaton,  
the wave lengths of the relevant perturbations become much greater
than the horizon distance during inflation and remain thereafter 
outside the horizon, uninfluenced by cosmic details, until they re-enter 
the horizon early in the matter era as density and matter perturbations.
 Among other things this neatly bypasses the complications of reheating     
 when the inflaton field is assumed to thermalize into a radiation 
 fluid. A parameter which is often used  to calculate the CMBRF, for the
 relevant small wave numbers $k$, is the metric curvature component 
 ${\cal R}_k$ which for these small $k$ is approximately given by
 $\zeta$ \cite{BST,LYTH} (with $\zeta \approx -{\cal R}$); in the simple 
 theory, having adiabatic dynamics of the physical quantities, this   
 is constant throughout the rest of inflation, reheating and the 
 radiation era.

 It has long been recognised \cite{MOLL,JGBDW} that that the
 constancy of $\zeta$ can alter in the presence of another scalar field,
 $\chi$. The question as to whether or not this occurs has to be addressed 
 for the details of each particular theory and there has been considerable
 interest \cite{BASS,LLMW,BGMK} in the effect of parametric resonance
 \cite{TRA,KLS1,etc,KLS2,K,KT,Boy} on this question, and on the CMBRF 
 generally. Quite a number of variations on this 
 theme have been studied \cite{KLS2,KT,GKLS,BASS}
 and some may well be significant for future directions. But
 basic scenarios are provided by the most investigated model
 where the interaction of the two fields is given by 
 $\frac{1}{2}g^2 \chi^2 \varphi^2$; parametric resonance may or may not 
 occur depending on the value of the constant $g^2$ and other parameters 
 of the theory. It is an object of this paper to investigate the above 
 questions for the $g^2 \chi^2 \varphi^2$ theories where $\chi$ has no
 initial classical component.

 A problem arising is treatment of the decoherence of the original 
 quantum perturbations and the consequent stochastic classical variables
 \cite{PS}. In most treatments the CMBRF are evaluated with the matter era 
 density and metric fluctuations (arising originally from the inflationary  
 era) as stochastic variations \footnote{Such fluctuations have also been
 derived, by-passing decoherence, by keeping the Heisenberg representation
 quantum operators throughout time, and evaluating the final correlation 
 functions in terms of the mode functions by taking (Heisenberg) vacuum 
 expection values\cite{ABB,GRI,HMM}.}

 In considering non-adiabaticity - whether or not accompanying
 parametric resonance - during reheating it is appropriate to 
 consider wave numbers of up to the order of the Hubble parameter.
 All waves of $\chi$ and and $\varphi_1$ arise originally as quantum
 perturbations, where $\varphi = \varphi_0 + \varphi_1$ with $\varphi_0$
 being the classical inflaton field. Then, treating all waves  
 similarly they eventually smoothly decohere, in the way shown by 
 Polarski and Starobinsky\cite{PS}, giving waves expressed by classical
 stochastic variables $e({\bf k})$ different for each value of the wave 
 number ${\bf k}$; the stage of decoherence depends on the number of the 
 field quanta of a given ${\bf k}$ as a funtion of time. We shall retain 
 the stochasticity throughout reheating and shall develop equations by 
 taking averages over stochastic variables. Typically this will give rise
 to terms such as $\langle\chi^2\rangle$ where $\langle ... \rangle$
 denotes the average; this is the equivalent of taking vacuum expectation 
 values in the quantum case, giving rise to a loop integral and being 
 sometimes in this context called the Hartree approximation.

  We have used stochastic variables, with the ensemble  
 averaging just mentioned; they have a smooth and well-defined 
 descent from the quantum operators \cite{PS}. A simple alternative, which  
 has been often used, is to insert at some time classical (non-stochastic) 
 fields for $\chi$ and $\varphi_1$. Besides its intuitive nature that 
 procedure has also the disadvantage of not being very well defined; so  
 for the present paper we have adopted this alternative excursion into  
 a more well defined treatment which has however its own approximation 
 in the above averaging.\footnote{The method of Khlebnikov and
 Tkachev \cite{KT} which is also based on stochastic fields is discussed 
 below in section III.}

 In our treatment there is no need for a perturbative expansion in
 $\chi$ and $\varphi_1$. In their occurence in the matter tensor 
 $T^{\mu}_{\nu}$ of the Einstein equations, or in
 the scalar equations, the non-linear terms pose no difficulties since we 
 have the averaging procedure and we get a set of simultaneous differential
 equations in the mode functions, as described in section III. For example
 $\langle\varphi_1^2\rangle$ quickly becomes comparable with $\varphi_0^2$   
 but this does not invalidate the equations. On the other hand we find it
 important to take account of the metric perturbation, $\psi$, and this 
 indeed we have to treat perturbatively and maintaining the perturbative
 validity imposes a strong limitation on the parameters of the theory.

 To achieve a state where the $\chi$ and $\varphi_1$ field densities  
 compare with the faded classical inflaton field density is not to complete
 reheating. The usual view is that reheating is completed when the 
 universe is dominated by a hydrodynamical fluid composed of relativistic
 particles. We go to such an era through the simplest 
 mechanism of  friction or dissipative terms with decay constants 
 $\Gamma_{\chi}$, $\Gamma_{\varphi}$ whose contribution to the $\chi$,
 $\varphi$ equations of motion respectively are proportional to 
 $\Gamma_{\chi} \chi'$, $\Gamma_{\varphi} \varphi'$ which operate effectively with 
 rapid oscillations of the fields. Since oscillations, for example 
 those associated with parametric resonance, are prominent features of the
 reheating process we expect such mechanisms or any dissipative decays, to 
 have a significant influence throughout reheating. Such
 dissipative processes are entropic.

 These features provide a model smooth physical connection of the reheat
 period to a later era with a radiation fluid. To embed the reheat period 
 precisely in a model cosmic history we need to specify the inflationary 
 era to provide initial conditions for the reheating. We choose power-law 
 inflation (partly for the convenience of an analytic solution) and this
 provides initial conditions both for the classical inflaton field 
 $\varphi_0$ and its quantum perturbations $\varphi_1$ with the associated
 metric perturbation. During the inflationary period the $\chi$ field 
 equation of motion has an important term $g^2 \chi \varphi_0^2$. For 
 any inflationary era in which $\varphi_0$ is large, O($m_{Planck}$),and
 non-oscillatory this term ensures extreme suppression of any initial 
 $\chi$ field. Thus the $\chi$ field, or equivalently the number of 
 $\chi$ particles, entering the reheat period is nearly zero. A smooth 
 junction with the reheat equations of motion then completes the model 
 embedding of the reheating in the assumed cosmic history.

 There now follows Section II in which we treat the relevant relations
 between quantum states and classical stochastic states. Sections III 
 and IV give the reheat equations, their initial conditions arising 
 from the inflationary model and some details of their solution. In
 Section V we discuss the results and we conclude with a general 
 summary in Section VI.

\section{STOCHASTIC FUNCTIONS IN REHEATING}\label{STOCHASTICITY}

Observational evidence \cite{SandB} on the cosmic microwave background 
fluctuations tends to support the theory that these are due to stochastic
perturbations of the cosmic density and metric arising from quantum
 perturbations of the early classical inflaton field, $\varphi_0(\tau)$
 \cite{LL}. These can be written as
  \begin{equation}\label{eq1}  \varphi_1({\bf x},\tau) = 
 \int \frac{d^{3}k}{(2\pi)^{\frac{3}{2}}} 
\lbrack c({\bf k})\varphi_{\bf k}(\tau)\exp(i{\bf k.x}) 
+ h.c. \rbrack 
\end{equation}
$\varphi_1({\bf x},\tau)$ is an operator in the Heisenberg picture, the
time dependence being wholly contained in the mode functions 
$\varphi_{\bf k}(\tau)$, and  $c(\bf k)$ is a time-independent 
quantum annihilation operator such that  
$\lbrack c(\bf k),c(\bf k')^{\dag}\rbrack=\delta^{3}({\bf k-k'})$.
 $\varphi_k(\tau)$ is a mode such that its $\tau$ dependence, 
for $\tau \rightarrow \infty$, is given by
  \begin{equation}\label{eq2}
 \varphi_k(\tau) \propto a^{-1}\exp{-ik\tau}
\end{equation}
We have adopted a conformal time metric whose unperturbed form, having
scale factor $a \equiv a(\tau)$, is
\begin{equation}\label{eq3} 
ds^{2} = -a^{2}d\tau^{2} + a^{2}\delta_{ij}dx^{i}dx^{j}.
\end{equation}

The most studied form of interaction has been with another scalar field 
$\chi$ with the interaction potential given by
\begin{equation}\label{eq4}
    V_{int}(\varphi,\chi) = \frac{1}{2} g^2 \varphi^2 \chi^2
\end{equation}
where $ \varphi =\varphi_0 + \varphi_1 $ and $\chi$ is in origin a 
quantum field. We adopt this as our representative of the basic scenario.
Like most other authors in this context we do not consider the case where 
$\chi$ has an aboriginal classical scalar field part; this would 
change the model to be more sophisticated like some variety of two-field
 inflation. Thus we write $\chi$ similarly to $\varphi_1$ as
  \begin{equation}\label{eq5}  \chi({\bf x},\tau) = 
 \int \frac{d^{3}k}{(2\pi)^{\frac{3}{2}}} 
\lbrack b({\bf k})\chi_{\bf k}(\tau)\exp(i{\bf k.x}) 
+ h.c. \rbrack
\end{equation}
where $b(\bf k)$ is a quantum annihilation operator ,
$ \lbrack b({\bf k}), b({\bf k}')^{\dag}\rbrack=\delta^{3}({\bf k-k'})$, and
 $\chi_{\bf k}(\tau)$ is a mode function such that its $\tau,k$ dependence, 
for $\tau \rightarrow \infty$, is given by
  \begin{equation}\label{eq6}
 \chi_{\bf k}(\tau) \propto a^{-1}\exp{-ik\tau}/\sqrt{2k}
\end{equation}
Thus we regard $\chi$ as originating like $ \varphi_1$. In the absence of
 a deeper theory, and in view of the seeming success in CMBR calculations
 of the theory of $ \varphi_1$, this seems a most reasonable ansatz.
 The $ \varphi_k(\tau)$ develop through the inflationary era and so do 
the $ \chi_k(\tau)$ mode functions.  

Now it is quite legitimate, for linear coupled equations of motion in 
the various quantum fields, to carry this development right through the 
reheat and radiation eras
  to the origin of the CMBR while evaluating the Heisenberg operators and 
finally taking vacuum expectation values of correlation functions 
\cite{AH,GRI,HMM}. In dealing with non-linear coupled equations it can be 
appropriate to make the transition to the more usual classical
 stochastic picture of the fields. Let us discuss this using as an 
example the scalar field equation for $\chi$ which is 
  \begin{equation}\label{eq7}
 \chi({\bf x},\tau)''+2(a'/a) \chi({\bf x},\tau)'-\nabla^2 \chi({\bf x},\tau)
+a^2 \lbrack M^2+g^2(\varphi_0(\tau)+\varphi_1({\bf x},\tau))^2 \rbrack
 \chi({\bf x},\tau) = 0 
\end{equation}
This is, from Eqs.(\ref{eq1},\ref{eq5}), a non-linear equation in the quantum 
operators
 where the non-linearity arises from the $g^2$ coupling of the operators
 in $\varphi_1$ with those in $\chi$. We eliminate the great quantum 
field theory difficulties thus arising by taking the Hartree approximation
 - discussed further below - on the $c,c^{\dag}$ in $\varphi_1^2$. That 
is we replace $(\varphi_0(\tau)+\varphi_1(x,\tau))^2$ by its vacuum
 expection value $\langle (\varphi_0(\tau)+\varphi_1(x,\tau))^2 \rangle_0
=  \varphi_0(\tau)^2+ \langle\varphi_1(x,\tau)^2\rangle_0$
  \begin{equation}\label{eq8}
 \langle\varphi_1(x,\tau)^2\rangle_0=(2\pi)^{-3}\int d^3k \varphi_k
\varphi_k^*
\end{equation}

In field theoretic terms this is a loop integral and as stated
in Eq.(\ref{eq17}) below the same integral form is maintained 
 when we go from quantum operators to stochastic variables. 
Now take the Fourier transform of the resulting approximation to
 Eq.(\ref{eq7}), leading to an equation linear in the creation and 
annihilation quantum operators $b_k$ and $b_k^{\dag}$.
The coefficients of $b_k$ and $b_k^{\dag}$ must each vanish giving two
equations which are complex conjugates. We could have reversed the order 
of the derivation and taken the vev after the selection of the coefficients. 
The resulting $\chi$ mode equation is 
  \begin{equation}\label{eq10}
\chi_k''+2(a'/a) \chi_k'
+(k^2+a^2\bar{M}^2)\chi_k = 0 
\end{equation} 
 \begin{equation}\label{eq9a}
\bar{M}^2(\tau)=
M^2+g^2\varphi_0(\tau)^2+g^2\langle\varphi_1({\bf x},\tau))^2\rangle_0
\end{equation}
It is important to note that $\chi$ is complex, as is the corresponding 
$\varphi_1$ in its equation, and the imaginary and the real parts are not
 constant multiples of each other. We have spelt out this derivation in 
 order to point up the similarities and the contrast to the semi-classical, 
 stochastic, case which now follows. 

Polarski and Starobinsky \cite{PS} have studied the transition from the 
quantum to the semi-classical case. For the Heisenberg picture which we 
have they find that when the mode functions grow large enough so that 
Planck's constant, $h$, can be neglected then the creation and 
annihilation operators may be replaced with appropriate Gaussian variables 
to form an equivalent stochastic field. In our formalism 
Eqs.(\ref{eq1},\ref{eq5}) are replaced by 
  \begin{equation}\label{eq11}  \varphi_1({\bf x},\tau) = 
 \int \frac{d^{3}k}{(2\pi)^{\frac{3}{2}}} 
\lbrack e({\bf k})\varphi_{\bf k}(\tau)\exp(i{\bf k.x}) 
+  e^*({\bf k})\varphi_{\bf k}^*(\tau)\exp(-i{\bf k.x}) \rbrack 
\end{equation}
  \begin{equation}\label{eq12}  \chi({\bf x},\tau) = 
 \int \frac{d^{3}k}{(2\pi)^{\frac{3}{2}}} 
\lbrack d(\bf k)\chi_{\bf k}(\tau)\exp(i{\bf k.x}) 
+  d^*(\bf k)\chi_{\bf k}^*(\tau)\exp(-i{\bf k.x}) \rbrack
\end{equation}
 $e(\bf k)$ and $d(\bf k)$ are time-independent separately 
 $\delta$-correlated Gaussian variables such that, where 
 $\langle .... \rangle$ denotes the average,
 \begin{equation}\label{eq13}
 \langle e({\bf k})e^*({\bf k}') \rangle=
 \langle d({\bf k})d^*({\bf k}') \rangle
 =\frac{1}{2} \delta^3({\bf k -k'});
 \end{equation}  
  \begin{equation}\label{eq14}
 \langle e({\bf k})e({\bf k}')\rangle=\langle d({\bf k})d({\bf k}')\rangle=0;
  \end{equation}
  \begin{equation}\label{eq15}
 \langle e({\bf k})d({\bf k}') \rangle=
\langle e({\bf k})d^*({\bf k}') \rangle=0,
  \end{equation}  
 that is the $e$ variables and the $d$ variables are uncorrelated.
 
 If now we insert Eqs.(\ref{eq11},\ref{eq12}) into Eq.(\ref{eq7}), take 
 the Fourier transform, and average over the Gaussian variables $e$ we 
 obtain, analogously to Eq.(\ref{eq10}),
  \begin{equation}\label{eq16}
\hat \chi_{\bf k}''+
2(a'/a)\hat \chi_{\bf k}'+(k^2 +a^2 \bar{M}^2) \hat \chi_{\bf k} = 0 
\end{equation}
 \begin{equation}\label{eq16a}
\hat \chi_{\bf k} \equiv d({\bf k})\chi_{\bf k}(\tau)+
d^*({\bf k})\chi_{\bf k}^{*}(\tau)
\end{equation} 
 where $\bar{M}^2$ is given by Eq.(\ref{eq9a}) since
 \begin{equation}\label{eq17}
 \langle\varphi_1^2\rangle=(2\pi)^{-3}\int d^3k \varphi_k\varphi_k^*=
 \langle\varphi_1^2\rangle_0
\end{equation}
The coefficients of the uncorrelated variables $(d-d^*)$ and $(d+d^*)$ 
 must each vanish separately and we can thus obtain
 \begin{equation}\label{eq18}
\chi_{\bf k}''+2(a'/a) \chi_{\bf k}'
+(k^2+a^2 \bar{M}^2)\chi_{\bf k} = 0 
\end{equation}
 where $\chi_k$ is complex and because of Eq.(\ref{eq17}) this is the 
 same as Eq.(\ref{eq10}).Thus we have a smooth transition from quantum
 to semi-classical or stochastic.

Polarski and Starobinsky have also shown that, in the semi-clasical 
 regime, the mode functions can be made real by a time independent phase
 transformation and they formulate the theory with somewhat different 
 Gaussian variables $e$ from those of Eqs.(\ref{eq14},\ref{eq15}). We
 can transform our classical stochastic formulation to that of 
 Polarski and Starobinsky:   

We can write Eq.(\ref{eq11}) as
 \begin{equation}\label{eq19}  \varphi_1({\bf x},\tau) = 
 \int \frac{d^{3}k}{(2\pi)^{\frac{3}{2}}} 
\lbrack e({\bf k})\varphi_{\bf k}(\tau) 
+  e^*(-{\bf k})\varphi_{-{\bf k}}^*(\tau) \rbrack \exp(i{\bf k.x}) 
\end{equation}
and we can time-independently transform our variables so that 
$\varphi_{\bf k}$ becomes real .Then using the fact that
 $\varphi_{\bf k}=\varphi_{-\bf k}$ we get
 \begin{equation}\label{eq20}  \varphi_1({\bf x},\tau) = 
 \int \frac{d^{3}k}{(2\pi)^{\frac{3}{2}}} 
\tilde e({\bf k})\varphi_{\bf k}(\tau) \exp(i{\bf k.x}) 
\end{equation}
where $\tilde e({\bf k}) \equiv  e({\bf k})+ e^*(-{\bf k})$. Thus
\begin{equation}\label{eq21}
\tilde e({\bf k})=\tilde e^*(-{\bf k}) 
\end{equation}
\begin{equation}\label{eq22}
\langle \tilde e({\bf k})\tilde e^*({\bf k}') \rangle=
\delta^3(\bf k -\bf k')
\end{equation}
where the last equation is derived using Eq.(\ref{eq13}).
 Eqs.(\ref{eq20}-\ref{eq22}) are those of Polarski and Starobinsky (except
 that we have the factor $a^{-1}(\tau)$ incorporated in our mode functions).

To summarise: using the Hartree approximation the scalar quantum and 
classical stochastic mode equations are the same, with complex mode 
functions, so there is a seamless join. In the classical stochastic regime
 a conversion to real mode functions can be made yielding completely real 
 equations, but these equations are invalid in the quantum regime. For 
this reason we prefer to use complex mode functions in reheat studies as
 not all modes may be classical stochastic through all the relevant period.

We note that if $\xi(k),\eta(k)$ are semi-classical mode functions, which 
 must be of dimension $(mass)^{-\frac{1}{2}}$, then we cannot write 
 equations such as
 $\xi(k)''+ .... +g^2\int \xi(k-k')\eta(k'-k'')\eta(k'')d^3k'd^3k'' = 0$ which 
 if valid would contain different and interesting rescattering effects. Apart 
 from being dimensionally inconsistent we would have had to take averages 
 over the stochastic variables in their derivation and this, as previously 
 above, would have eliminated all such terms. Of course such equations are 
 perfectly possible if $\xi,\eta$ are fully classical mode functions of 
 dimension $(mass)^{-2}$ and indeed have often been written down in the 
 resonant reheating context. But to obtain such such equations, starting
 from a quantum origin of the modes as we do, requires specific
 transformations and we are not aware of any suggestions for these for 
 $k \not =0$. 
 Naturally if $\xi(k)=d(k)\chi_k ,\eta(k)=c(k)\varphi_k$ then such equations 
 are valid, but the physical solution involves solving a statistical 
 ensemble of equations, got by sampling $d,e$ and finally averaging - a 
 seemingly very lengthy procedure. The method of Khlebnikov and Tkachev 
 \cite{KT}, mentioned also below, seems to provide at least a partial 
 and more achievable way of implementing the stochasticity in the 
 semi-classical case without resort to the Hartree approximation. 
 However having a system of 7 simultaneous equations with complex mode 
 functions we shall implement the quantum or semi-classical equations 
 which we have illustrated above.

As a final comment we note that it is dimensionally obviously inconsistent
to write such coupled classical equations as the above and loop integrals 
such as (\ref{eq17}) with the same mode functions.

\section{THE REHEATING EQUATIONS}\label{III}

We now formulate the equations of motion for the reheat period. The
 Lagrangian for the scalar fields is
 \begin{equation}\label{eq23}
L=\int d^4x\sqrt{-g}\lbrack\frac{1}{2}\varphi^{,\alpha}\varphi_{,\alpha}+
\frac{1}{2}\chi^{,\alpha}\chi_{,\alpha}-V(\varphi)-V(\chi)-
V_{int}(\varphi,\chi)\rbrack 
\end{equation}
 where $\varphi=\varphi_0 + \varphi_1$ and $V_{int}$ is given by 
 Eq.(\ref{eq4}).
  
In the perturbed metric, $g_{\alpha,\beta}$, we use the longitudinal 
 gauge for the perturbation
\begin{equation}\label{eq24}
ds^2 = a(\tau)^2(1+2\phi)d\tau^2-a(\tau)^2(1-2\psi)\delta_{ij}dx^idx^j 
\end{equation}
Unusually, the scale factor is calculated using the full reheating dynamics
 as below, Eq.(\ref{eq27}). Unlike other treatments, excepting Bassett et al. 
 \cite{BASS}, we include the metric perturbations as dynamical variables in 
 the coupled equations of motion. In this physical system, including the 
 hydrodynamical variables introduced below, there are no space-space 
 non-diagonal components in the energy-momentum tensor. Hence $\phi=\psi$
 \cite{MUK} and we denote this metric variable by $\psi$.

 We also wish to take acount of actual reheating in which the $\varphi$ and 
 the produced $\chi$ particles, which may be massive \cite{KRT},are 
 replaced by a thermal gas of relativistic particles, giving the radiation 
 era. We make no attempt to imagine details of this process but adopt the 
 simple friction mechanism which puts all these details into a black box.
 The use of this mechanism in general cicumstances has been criticized, 
 it being claimed that it is only appropriate when the decaying field is 
 undergoing fast oscillations \cite{K}. In the calculations that we shall 
 describe it is in these circumstances that the main production of a 
 relativistic gas occurs. We describe this gas by the usual hydrodynamical 
 variables $\rho$, the density, and $p$ the pressure and in the equations 
 of motion which follow we take that equation of state, $p=\rho/3$, which 
 is appropriate for relativistic particles.

We divide the hydrodynamical density into two components 
\begin{equation}\label{eq25}
  \rho({\bf x},\tau) = \rho_0(\tau) + \rho_1({\bf x},\tau) 
\end{equation} 
where $\rho_0(\tau)$ is the homogeneous background density and 
 $\rho_1({\bf x},\tau)$ is the inhomogeneous part of it; the notation 
 $\delta \rho \equiv \rho_1 /\rho_0$ is the usual usage. As we shall see
 in the equations of motion $\rho_1$ mainly arises from the inhomogeneous 
 part $\varphi_1$ of the inflaton field in accord with the usual theory of 
 the cosmic microwave background fluctuations.

We shall now formulate the equations of motion, being those Einstein 
 equations and scalar wave equations arising from the Lagrangian, 
 Eq.(\ref{eq23}), and metric, Eq.(\ref{eq24}), with the addition of the 
 hydrodynamical terms. The energy-momentum tensor is
 \begin{equation}\label{eq26}
T^{\mu}_{\nu} =\varphi^{,\mu}\varphi_{,\nu}+\chi^{,\mu}\chi_{,\nu}-
\lbrack\frac{1}{2}\varphi^{,\alpha}\varphi_{,\alpha}+
\frac{1}{2}\chi^{,\alpha}\chi_{,\alpha}+
V(\varphi)+V(\chi)+V_{int}-p\rbrack \delta^{\mu}_{\nu}+
(\rho + p)u^{\mu}u_{\nu}
\end{equation}  
 where $u^{\mu}$ is the fluid 4-velocity.

First we write 3 spatially homogeneous ,or background, equations taking 
 the average over the Gaussian variables as in Eqs(\ref{eq13}-\ref{eq17}).
 We use the notation $'$ for $d/d\tau$ and we take the individual field 
 potentials in reheat to be:
\begin{equation}\label{eq26a}
 V(\varphi)=\frac{1}{2}m^2\varphi^2 ; V(\chi)=\frac{1}{2}M^2\chi^2. 
\end{equation} 
After Eq.(\ref{eq32}) below it is noted that $\varphi_1$ and $\psi$ have 
the same stochastic variables. Thus
defining the density and pressure homogeneous parts of the 
 energy-momentum tensor as $\rho_T(\tau) \equiv -\langle T^0_0 \rangle$ 
 and $p_T(\tau)\delta^i_j \equiv \langle T^i_j \rangle$ we find, after 
 ensemble averaging, that
\begin{equation}\label{eq26b}
 \rho_T(\tau) = \frac{1}{2 a^2} \lbrack \eta 
+a^2\bar{m}^2(\varphi_0^2+\langle \varphi_1^2 \rangle)
+a^2M^2 \langle \chi^2 \rangle
+\langle \varphi_{1,i}^2 \rangle)+\langle \chi_{,i}^2 \rangle
 -4\varphi'_0\langle\varphi_1'\psi\rangle\rbrack+\rho_0 ,
\end{equation} 
\begin{equation}\label{eq26c}
 p_T(\tau)=\frac{1}{2 a^2} \lbrack \eta
-a^2\bar{m}^2(\varphi_0^2+\langle \varphi_1^2 \rangle)
-a^2M^2 \langle \chi^2 \rangle
-\langle \varphi_{1,i}^2 \rangle)/3-\langle \chi_{,i}^2 \rangle/3
 -4\varphi'_0\langle\varphi_1'\psi\rangle\rbrack+\rho_0/3 , 
\end{equation} 
\begin{equation}\label{eq26d}
 \eta =  
 \varphi_0'^2+\langle\varphi_1'^2\rangle+\langle \chi'^2 \rangle
\end{equation} 
\begin{equation}\label{eq28}
\bar{m}^2 \equiv m^2+g^2\langle \chi^2 \rangle.
\end{equation}
and we have the Friedmann equation
\begin{equation}\label{eq27}
(a'/a)^2= \frac{8\pi G}{3}a^2\rho_T(\tau).
\end{equation}

As in Eq.(\ref{eq17}) the averages are independent of ${\bf x}$ ensuring 
 the spatially homogeneity of the R.H.S. of Eq.(\ref{eq27}). We also see
 from Eq.(\ref{eq17}) that Eq.(\ref{eq27}) is equally valid in the quantum 
 regime since the formalism ensures that the vev of the quantum operators 
 equals the average over the stochastic variables.

The other two spatially homogeneous equations are for 
 $\varphi_0$ and $\rho_0$. Here we introduce the arbitrary decay constants 
 $\Gamma_{\varphi}$ and $\Gamma_{\chi}$ which serve to create the 
 hydrodynamical radiation gas specified by $\rho=\rho_0+\rho_1, p=\rho/3$.
 In Appendix A we show how these are introduced in a way consistent with
 the Bianchi identities and how the equations for 
 $\rho_0,\rho_1, \varphi_0,\varphi_1,$ and $\chi$ are subsequently 
 derived.
\begin{equation}\label{eq29}
\varphi_0''+2(a'/a)\varphi_0'+a^2 \bar {m}^2 \varphi_0
-4\langle \psi'\varphi_1' \rangle-4\langle \psi\nabla^2\varphi_1\rangle
+2a^2\bar{m}^2 \langle \psi\varphi_1 \rangle=
-a\Gamma_{\varphi}(\varphi_0'+2\langle \psi\varphi_1' \rangle)
\end{equation}

We note the combined reaction of the inhomogeneous fields $\varphi_1$ and 
 $\psi$ on the homogeneous inflaton field through the ensemble averaging,
 similarly to that in Eqs. (\ref{eq26b}) and (\ref{eq26c}).

\begin{equation}\label{eq30}
\rho_0'+4(a'/a)\rho_0=
a^{-1}\Gamma_{\varphi}(\varphi_0'^2+\langle \varphi_1'^2 \rangle)+
a^{-1}\Gamma_{\chi}\langle \chi'^2 \rangle
-2a^{-2}\varphi_0'\langle \psi\nabla^2\varphi_1\rangle
+4\langle \psi'\rho_1\rangle
\end{equation}
There remain the 4 spatially non-homogeneous equations which we write in 
the $k$-component form. For $\chi$ and $\varphi_1$ these components are 
specified as the complex mode functions $\chi_k$ and $\varphi_k$ of 
Eqs.(\ref{eq5},\ref{eq11},\ref{eq12}). Their wave number dependence, given 
 by the succeeding equations, is only on $k \equiv \left|{\bf k}\right|$ . 
\begin{equation}\label{eq31}
\chi_k''+2(a'/a) \chi_k'
+(k^2+a^2\bar{M}^2)\chi_k = -a\Gamma_{\chi}\chi_{k}' 
\end{equation}
where $\bar{M}^2$ is a function of $\tau$ given by Eqs.(\ref{eq9a},\ref{eq17}).
\begin{equation}\label{eq32}
\varphi_k''+2(a'/a)\varphi_k'+(k^2+a^2 \bar {m}^2) \varphi_k -
4\varphi_0'\psi_{k}'+
2a^2\bar{m}^2\varphi_0\psi_k=
 -a\Gamma_{\varphi}(\varphi_k'+2\varphi_0'\psi_{k})
\end{equation}                  
Though not indicated these equations actually hold for each separate value of
 ${\bf k}$; thus consistency of Eq.(\ref{eq32}) indicates that $\psi_{\bf k}$, 
 the mode function of the metric perturbation should be associated with the 
 the same Gaussian operators (or,in the quantum regime, with the same 
 quantum operators) as $\varphi_{\bf k}$ in Eq.(\ref{eq11}). Thus
 \begin{equation}\label{eq33}  \psi({\bf x},\tau) = 
 \int \frac{d^{3}k}{(2\pi)^{\frac{3}{2}}} 
\lbrack e({\bf k})\psi_{\bf k}(\tau)\exp(i{\bf k.x}) 
+  e^*({\bf k})\psi_{\bf k}^*(\tau)\exp(-i{\bf k.x}) \rbrack 
\end{equation}          
where since $\psi({\bf x},\tau)$ is dimensionless $\psi_{\bf k}$ has 
 dimension $(mass)^{-\frac{3}{2}}$. This associates the metric perturbation 
 with the inhomogeneous part of the inflaton field without assigning 
 priority to  either. But the stochastic variables of $\chi_{\bf k}$ are 
 independent. Thus no terms in $\psi_{\bf k}$ appear in Eq.(\ref{eq31});
 they are forbidden through ensemble averaging.. 

The mode equations for $\psi$ and $\rho_1$ are

\begin{equation}\label{eq34}
\psi_k''+(3\psi_k'+(a'/a)\psi_k)(a'/a) = 
 \frac{8\pi G}{3}a^2(\rho_T + 3p_T)\psi_k + 4\pi G a^2 \delta p_k
\end{equation}
 where

\begin{equation}\label{eq34a}
 \delta p_k = \frac{1}{a^2}\lbrack -\xi\psi_k + \varphi_0'\varphi_k'
-2\varphi_k'\langle \psi\varphi_1' \rangle
 -a^2\bar{m}^2\varphi_0\varphi_k \rbrack+\rho_k/3
\end{equation}
with
\begin{equation}\label{eq35}
\xi \equiv \varphi_0'^2+\langle\varphi_1'^2\rangle+\langle\chi'^2\rangle+
\langle\varphi_{1,i}\varphi_{1,i}\rangle/3+\langle\chi_{,i}\chi_{,i}\rangle/3.
\end{equation}
										
 \begin{equation}\label{eq36}
\rho_k'+4(a'/a)\rho_k -4\rho_0\psi_k' =-\frac{k^2}{a^2} \lbrack 
(4\pi G)^{-1}(\psi_k'+(a'/a)\psi_k)-
(\varphi_0'+2\langle \psi\varphi_1' \rangle)\varphi_k \rbrack
+2\Gamma_{\varphi}\varphi'_0\varphi'_{k}
\end{equation}
Eqs.(\ref{eq34})-(\ref{eq36}) can be deduced by extending the space-time
and space-space gauge invariant Einstein equations given by 
Mukhanov et al.\cite{MUK}; the term in square brackets in Eq.(\ref{eq36})
 arises from the validity of the space-time equation and vanishes if
 $\rho_0 = 0$. 

 From the structure of the above equations the Fourier transform 
of $\rho_1({\bf x},t)$ has the same Gaussian variables, $e({\bf k})$, as
have $\varphi_1({\bf x},t)$ and $\psi({\bf x},t)$, $\rho_{\bf  k}$ 
having dimension $(mass)^{\frac{5}{2}}$; the quantum operators and
consequent Gaussian variables of the $\chi$ field are independent
of the others. This point also has consequences when we 
 use vacuum expectation values or averages over the stochastic variables
 as noted above in connection with Eq.(\ref{eq31}).

A number of points arise on our equations of motion in the context of
the large body of work on parametric resonance in reheating since the
subject was introduced by Traschen and Brandenberger \cite{TRA}:

 {\it i. Scale factor.}
 We find the scale factor $a(\tau)$ by Eq.(\ref{eq27}) this being the 
 appropriate calculation which includes all the relevant dynamical 
 variables, and thus the influence of all the scalar fields. Generally
 an approximate ansatz for $a$ has been used in other paper, an 
 exception being the work of Boyanovsky et al.\cite{Boy}. Our
 equation makes a difference to the form of $a$ but this change does not 
 seem to have a critical influence.

 {\it ii. The loop integrals (Hartree approximation).} 
  For quadratic field forms, in the classical stochastic regime, we have 
 used the Gaussian average which is a smooth continuation from the quantum 
 field theory vacuum expectation value. This has been called the 'Hartree 
 approximation' and has been much used in parametric resonance work. (We 
 give some details of our evaluation in Appendix B.) We note that, for 
 example, the term $g^2\varphi_0^2 \langle \chi^2 \rangle$ corresponds  
 to a first order single loop calculation in quantum field theory as 
 discussed by  Kofman et al.\cite{KLS2}. They have compared the magnitude 
 of this with that of an approximate evaluation of the single loop second 
 order in $g^2$ term, and have concluded that this could be of comparable 
  magnitude for some resonance modes they studied. In common with Kofman 
 et al. and other authors we have not included this difficult, though 
 maybe significant, correction. To include suchlike corrections we would have 
 to add the results of the appropriate field theory calculations to 
 Eq.(\ref{eq23}), thus manufacturing an effective Lagrangian. 

 As discussed in the last paragraph of section \ref{STOCHASTICITY} our 
 formalism does not include rescattering corrections of the form 
 $\int \xi(k-k')\eta(k'-k'')\eta(k'')d^3k'd^3k''$ as our regime is 
 stochastic. Other authors \cite{KLS2,BASS} have written down, though perhaps
 not fully implemented , equations with such terms which seem to require 
 the fields  to be already fully classical, in the sense noted in section 
 \ref{STOCHASTICITY} together with a question on their provenance. We have 
 found that our emphasis on the Hartree approximation is closely related to 
 the viewpoint of Boyanovsky et al. \cite{Boy}.

 {\it iii.The metric perturbation.}
 The coupled equations include the metric perturbation, $\psi$. As noted 
 above its quantum or stochastic variables are the same as those of 
 $\varphi_1$ whose equation of motion (\ref{eq32}) has an important 
 coupling to $\psi$  arising from the metric interaction with
 $\varphi_0$. The equation of motion (\ref{eq34}) of $\psi$ with 
 Eqs.(\ref{eq34a},\ref{eq35}) show its coupling to all the other fields.
 In the derivation of the coupled equations $\psi$ is treated as a 
 small perturbation to the metric, this being in contrast to $\varphi$
  and $\chi$ which are not treated perturbatively. So the validity of
  this has to be maintained throughout reheating and we find this 
  gives significant restrictions on the parameters of the theory.
  Much work involving solving equations of motion in this context 
  has ignored the metric perturbation but Bassett et al.\cite {BASS} 
  have discussed and emphasized its importance.

 {\it iv. Stochasticity.}
  Khlebnikov and Tkachev \cite{KT}, as mentioned also in section 
 \ref{STOCHASTICITY}, have an entirely different approach to the 
 equations of motion, which they solve on an x-space lattice. Their method 
 takes account of the stochastic nature of the variables, as exhibited for 
 example in Eq.(\ref{eq11}), by taking appropriate initial conditions. For
 each discretized $k$-value they take a random value (chosen from an 
 appropriate distribution) of for example the variable 
 $e({\bf k})\varphi_{\bf k}(\tau)$ of Eq.(\ref{eq11}). The initial value of
 $\varphi({\bf x},\tau)$ is then got by using the discretized 
 Eq.(\ref{eq11}) to sum over all those random values. Thus though only one
 value is used for each ${\bf k}$ each ${\bf x}$ contains a large random 
 sample.After solution in $x$-space the variables can be re-Fourierized to
 get the ${\bf k}$ components. Because of the non-linearity of the $x$-space 
 equations this mixes the initial ${\bf k}$ components and is not equivalent 
 to solving the ${\bf k}$ equations directly. We note that the fields 
 must already have passed through the quantum stage into the classical 
 stochastic phase; the authors do not include the metric perturbation .

\section{Solving the Equations}\label{IV}

\subsection{The Microwave Background Fluctuations}\label{IVA}
The present inflationary theory of the origins of the cosmic microwave 
background fluctuations has so far been successful. In its simplest form 
a perturbation, $\delta\varphi_{k}(t)$ of quantum origin, in the inflaton 
field developes in the inflationary era in association with the curvature 
perturbation, ${\cal R}_{k}(t)$. At a certain time, $t_*$, in the 
inflationary era before reheat begins, the modes of relevance are such that
$k/a$ becomes equal to, and then rapidly less than, $\dot a/a \equiv H$. Since 
then $\lambda_{physical}=2\pi a/k \gg H^{-1}$ the latter being, in this 
context, the Hubble radius and this is sometimes expressed  
as the perturbations having passed outside the horizon. At around the 
beginning of the period of matter domination $H^{-1}$ again becomes greater 
than $\lambda_{physical}$. 
 ${\cal R}_{k}(t_*)=-\lbrack H\delta\varphi_{k}/\dot \varphi \rbrack_{t=t_*}$ 
and, with certain assumptions in the simple model such as adiabaticity, 
 ${\cal R}_{k}$ remains the same during that whole period when
 $\lambda_{physical} \gg H^{-1}$ and equivalent \cite{LYTH} to the  
 parameter $\zeta$ \cite{BST,MUK,LYTH,BARDN}:
 \begin{equation}\label{eq37}
\zeta_k = \frac{2}{3}(H^{-1} \dot\Phi_k + \Phi_k)/(1+w) + \Phi_k = 
 -{\cal R}_{k} 
\end{equation}
where $\Phi_k$ is the gauge invariant metric perturbation (equal to the 
longitudinal gauge metric perturbation $\psi_k$ of Eq.(\ref {eq24})) and
$w=p/\rho$ being the total pressure to density ratio. This scenario 
relates the CMBRF rather directly to the inflationary period bypassing 
reheating complications.

Now the model which we are studying has an extra scalar field, with an 
interaction $g^2 \chi^2 \varphi^2$ known to be capable of giving rise to 
parametric resonance (preheating \cite{KLS1,KLS2}) in reheating .  
We consider how far and under what conditions, the constancy of $\zeta$
for $\lambda_{physical} \gg H^{-1}$ remains valid. For this purpose 
we investigate the conclusion of reheating by particle decay, as mediated 
by the $\Gamma_{\varphi}$ and $\Gamma_{chi}$ parameters and for 
comparison we also look at cases where these are zero.

 In what follows we shall largely be engaged with modes which have 
 wavelengths which do {\it not} become larger than $H^{-1}$ before reheat. 
 So we should have an estimate of the magnitude of all relevant
 modes at the beginning of reheat and this will be dealt with in the 
 following subsection. 

\subsection{Initial Conditions at Reheat}\label{IVB}

\subsubsection{$\chi$-field}\label{IVB1}

Whatever we may postulate in detail for the single field potentials
 $V(\varphi),V(\chi)$ through the inflationary and reheat periods, 
it is a hypothesis of our model that the $g^2\varphi^2\chi^2$ is the 
significant interaction term in the Lagrangian. This has some immediate 
consequences for the magnitude of $\chi$ at the (fuzzy) interface between 
inflation and reheat. A simplified version of the mode equation, 
Eq.(\ref{eq31}) with Eq.(\ref{eq9a}), in the inflationary era is 
 \begin{equation}\label{eq38}
(a\chi_k)''+[k^2+a^2 M^2+g^2 a^2 \varphi_0^2-a''/a](a\chi_k)=0
 \end{equation}
and $\varphi_0^2$ is greater than, or of the order of, $m_{Pl}^2$ 
through most of the inflationary era. Thus the term in square brackets
 is large and positive resulting in a quasi-periodic type solution 
 for $a\chi_k$, which has a primordial value given by Eq.(\ref{eq6}).
 The many efold increase of $a$ during the inflationary era indicates 
an exceedingly small value for $\chi_k$ at the beginning of reheat 
\cite{BassV3,JS}, which may be as little as $10^{-50}$ of its 
 initial value. This is an important qualitative feature. Firstly 
for such small mode functions in the beginning of reheat the transition 
to classical stochastic functions cannot yet be made and 
the quantum complex formalism should be retained. We have taken
the initial value of $\chi$ to be $10^{-n}m_{Planck}^{-1/2}$ where 
 $n$ is of the order of 30.

\subsubsection{Background metric and the classical 
 inflaton field}\label{IVB2}

For the reasons just outlined we shall assume that it is valid to
 neglect any influence of the $\chi$-field during inflation on  
 the classical background field $\varphi_0$ and on the quantum 
 or stochastic part $\varphi_1$. So we shall work with the 
 standard single field inflation formalism to formulate conditions 
 at the beginning of reheat for both $\varphi_0$ and $\varphi_1$. It 
 will prove convenient, as we shall see in {\it 4} below, to have
 a potential, $V(\varphi)$, of different form in the inflation era from 
 that in the reheat era. In a perfect physical model these should be 
 joined by a smooth transition form but as we show we can substitute 
 a sudden (phase change) transition for the purpose of finding the 
 initial conditions of reheat. First we deal with the classical 
 background field which we denote just by $\varphi$ in the rest of
 this section.

 At the boundary between inflation and reheat we can start the reheat 
 era with the same values of $\varphi$ and $\varphi'$ as those at the 
 end of inflation and we can likewise impose continuity of the potential 
 $V$ (for an example see {\it 4} below); of course we also take $a$ 
 continuous. It follows that $a'$ is continuous since
\begin{equation}\label{IVB2.1}
 \alpha^2 \equiv (a'/a)^2= \frac{8\pi G}{3} 
\lbrack\frac{1}{2}\varphi'^2+a^{2}V \rbrack
\end{equation}
 We can show that $\alpha'$ is continuous as follows.
 \begin{equation}\label{IVB2.2}
 2\alpha\alpha' = \frac{8\pi G}{3} \lbrack \varphi'\varphi''+
 a^{2}\varphi'(dV/d\varphi) + 2\alpha a^2 V\rbrack
\end{equation}
 Since 
\begin{equation}\label{IVB2.3}
 \varphi'' + 2\alpha\varphi' + a^2 dV/d\varphi = 0
\end{equation}
 it follows that
\begin{equation}\label{IVB2.4}
 \gamma \equiv 1-\frac{\alpha'}{\alpha^2} = 4\pi G(\varphi'/\alpha)^2 
\end{equation}
 where $\gamma$, thus defined, is a quantity we shall need in the next 
 section.

 Thus from the imposed conditions at the beginning of reheat it follows 
 that $\alpha'$, or equivalently $a''$, is continuous at the boundary.
 Thus so is the background curvature, in addition to the metric.

\subsubsection{$\varphi_1$ and the metric perturbation}\label{IVB3}

Having a satisfactory background continuation we can now deal with the
continuation of the quantum perturbations of the inflationary  
era. Deruelle and Mukhanov \cite{DER} 
have used the Lichnerowitz curvature conditions \cite{LIC} to give the 
necessary continuity conditions on the metric perturbation. These can be 
summarised as follows\cite{HMM,MHM}: $\psi$, the metric perturbation and 
$\Gamma$ should be continuous across the boundary where 
\begin{equation}\label{IVB3.1}
 \Gamma \equiv (\psi'/\alpha + \psi - \nabla^2 \psi/3\alpha^2)/\gamma . 
\end{equation}
Since $\gamma$ has just been found to be continuous the last condition 
 can be reduced to the continuity of $\psi' + \alpha\psi$, which in 
 turn using the time-space Einstein equation \cite{MUK}
\begin{equation}\label{IVB3.2}
 \psi' + \alpha\psi = 4\pi G\varphi_0' \varphi_1
 \end{equation}
 can be reduced to the continuity of $\varphi_1$.

Thus from the Lichnerowitz conditions one finds that the values of 
 the mode functions for the metric perturbation $\psi$ and for the 
 non-homogeneous part, $\varphi_1$, of the  inflaton field
 at the beginning of reheating after the phase change 
 are the same as those at the end of the inflation era; $\psi'$, given 
 by Eq.(\ref{IVB3.2}), is also continuous. $\varphi_1'$ does not need
 to be continuous and in fact is not. Its value can be found in terms
 of the continuous fields by using the time-time Einstein equation
 at the beginning of reheating (which differs from that at the
 end of inflation). At this very beginning of reheat, as at the end of
 inflation, we validly use linear perturbation theory in $\varphi_1$ and
 we can check that the perturbations are adiabatic. This means that we
 expect $\zeta$ to be constant in the first part of the reheat period, 
 which is what we find as discussed below in section \ref{VC}.

 The values of the above quantities with $\varphi_0,\varphi'_0$ and 
 $\alpha \equiv (a'/a)$, plus the input of the very small $\chi,\chi'$, 
 provide all that is needed for the initiation of the reheat equations.

\subsubsection{Input from power-law inflation}\label{IVB4}

 Rather than making some plausible input data from a generality of 
 of inflationary calculations we have chosen to use a specific 
 inflationary model. This is power-law inflation with an exponential 
 potential: $V =  Uexp(-\lambda\varphi)$ where $U$ is a constant.
 In this there are 2 parameters free to be chosen.
 These are the specific power of the inflationary scale factor $a$,
 and the magnitude of the classical
 scalar field $\varphi_0$ at the junction of inflation and  
 reheat, which we define as where we begin to use the reheat equations 
 of section \ref{III}.

 An advantage of this power-law inflation is that the solutions 
 are analytically expressible \cite{MHM}.
 For the scale factor $a$ and the inflaton field $\varphi_0$
\begin{equation}\label{IVB4.1}
 a \propto (\tau_i - \tau)^p , \varphi_0'= -\alpha\lambda/\kappa,
\end{equation}
where $\tau_i > \tau$(end inflation) is a constant and 
 $p=2\kappa/(\lambda^2-2\kappa)$. With $\lambda^2/\kappa \ll 2$, $p$ is near 
to $-1$ and there is power-law inflation. The imposed continuity of $V$ at
the boundary gives $Uexp(-\lambda \varphi_0) = \frac{1}{2}m^2 \varphi_0^2$;
 thus, given $m,\lambda$, we can arbitrarily specify $\varphi_0$ at the 
 boundary and this condition then merely specifies $U$. This triviality in 
 making an arbitrary choice of $\varphi_0$ at the beginning of reheat is 
 useful.

Also, in the inflation era, the perturbation $\varphi_1({\bf x},\tau)$ and 
the perturbation $\psi({\bf x},\tau)$ to the metric can be expressed in 
 terms of a quantum field function 
\begin{equation}\label{IVB4.2}
\mu({\bf x},\tau) = 
a^{-1}\int \frac{d^3k}{(2\pi)^{3/2}\sqrt{2k}}
\lbrack a({\bf k})\mu_{\bf k}(\tau)\exp(i{\bf k.x}) + h.c. \rbrack
\end{equation}
whose $k$-modes satisfy
\begin{equation}\label{IVB4.3}
\mu_k'' + (k^2-a''/a)\mu_k = 0
\end{equation}
This function emerges in the synchronous gauge, a derivation due to 
Grischuk \cite{GRI,HMM};one can convert the resulting formulae to our
longitudinal gauge quantities by well known techniques \cite{MUK,HMM}.
These yield
\begin{equation}\label{IVB4.4}
\psi_k = a^{-1}\sqrt{\frac{4\pi G}{2k}}
\alpha\sqrt{\left|\gamma\right|}(\mu'_k-\alpha\mu_k)/k^2
\end{equation}
\begin{equation}\label{IVB4.5}
\varphi_k = a^{-1}(2k)^{-1/2}[\mu + \alpha\gamma (\mu'-\alpha\mu)/k^2]
\end{equation}
where $\gamma \equiv 1-\alpha'/\alpha^2=(p+1)/p$.
Since $a''/a=\frac{p+1}{p} (\tau_i - \tau)^{-2}$, Eq.(\ref{IVB4.3}) can be 
expressed as a Bessels equation in terms of
\begin{equation}\label{IVB4.6}
y \equiv  k(\tau_i-\tau)=\frac{k}{aH}\left|p\right|
\end{equation}
and the solutions which satisfy the asymptotic conditions for a scalar field
as in Eq.(\ref{eq6}) are
\begin{equation}\label{IVB4.7}
\mu_{k}(y) = \sqrt{\frac{\pi y}{2}}(J_{n} - iY_{n})
\exp\lbrack-i(\frac{1}{2}n\pi + \frac{1}{4}\pi)\rbrack
\end{equation}
where $n=1/2-p$.

For $k/a \ll H$ (many times over fulfilled by the $k$ relevant  
to the CMBRF observations)
$\mu_{k}(y) \propto y^p \propto k^p$ and if this $k$-dependence passes 
unaltered through the reheat era into the radiation era it yields a power-law
spectrum which scales as $k^{n-1}$ where $n$, the spectral index, is given by
$n=2p+3$.

  Besides $\varphi_0$ and the power $p$ there is one more parameter that 
  determines the input from inflation and that is the $m$ of the inflaton 
  field potential in reheat, $V(\varphi)=\frac{1}{2} m^2 \varphi^2$. This 
 is because it is necessary to know the 
 value of $y$, defined by Eq.(\ref{IVB4.6}), in order to determine
 the size of the perturbations. From the conditions of continuity of 
 section \ref{IVB2} it follows that at the junction of inflation and 
 reheat  
 \begin{equation}\label{VA1.1}
     H = \sqrt{4\pi G}m\varphi_0 \sqrt{p/(2p-1)}       
\end{equation}
 and one can use this in Eq.(\ref{IVB4.6}) to give
 \begin{equation}\label{VA1.2}
     y = \frac{k}{aH}\left|p\right|       
\end{equation}
 To summarize: $m\varphi_0$ and $p$
 determine the input of $\psi$ and $\varphi_1$ to the reheat era and 
 the reheating equations of Section III determine subsequent developments.

\section{Results}\label{V}

Our interest is in the qualitative features of reheat in a model
 compatible with known observations rather than in making precise
 comparisons with data. So we are interested in values of the parameters 
 that roughly supply  the usual requirements of inflation and the 
 magnitude of the CMBR fluctuations.

Unless otherwise stated the results we quote are for the values
 $m = 10^{-7} , \varphi_0 = 0.3 , p = -1.1, M/m=0.02$.Throughout we quote 
 dimensionful results and parameters in units such that 
 $\hbar = c = G =1$. 

  The critical parameters are the coupling constant $g^2$ and the 
  frictional decay constants $\Gamma_{\chi}$and $\Gamma_{\varphi}$ of the
  $\chi$ and $\varphi$.
 The quartic coupling constant $g^2$ governs the number of $\chi$ 
 produced and in particular the parametric resonance. We have considered 
 values in the range given by $20,000 > g/m > 1000$. For smaller values
 there is no appreciable particle production. The range of consideration 
 for the frictional decay constants was 
 $0.1 \ge \Gamma_{\chi}/m \ge 0$ and $0.005 \ge \Gamma_{\varphi}/m \ge 0$.
 Larger values of $\Gamma_{\chi}$ tend to suppress parametric
 resonance while larger values of $\Gamma_{\varphi}$ may be unrealistic 
 in the context of the development of the classical inflaton field.
 We have by no means done a complete scan over this parameter range
 but in each particular case we have looked at (i) the existence or not 
 of parametric resonance (ii) the development of the metric inhomogeneity
 $\psi$ and also $\zeta$ (iii) the post reheating state; and other
 significant features.  To illustrate significant features the figures show 
 a case with  parametric resonance in the production of $\chi$ particles and 
 $\Gamma_{\chi}/m = 0.001$, $\Gamma_{\varphi}/m=0.0004$ so that no 
 $\chi$ particles are finally left. 
 If both $\Gamma 's$ are zero then reheating
 (as opposed to preheating) does not happen. But it is interesting to 
 inspect this case (also shown in the figures) 
 to look at what the possible variation of $\zeta$ might be
 if the decay of the $\chi$ and $\varphi$ particles is significantly slower.

\subsection{The Metric Perturbation,$\psi$}\label{VD}

As noted in section \ref{III} the metric is given by
$ds^2 = a(\tau)^2(1+2\psi(x,\tau))d\tau^2-
a(\tau)^2(1-2\psi(x,\tau))\delta_{ij}dx^idx^j$ .From the end of reheat 
$\psi$ develops through the radiation and matter eras so that the magnitude 
of the components $\psi(k,\tau)$ for very small $k$ is a determinant of 
the microwave background perturbation.

\subsubsection{The magnitude of $\psi$}\label{VD1}
The validity of the perturbation formalism depends on $\psi(x,\tau)$ being 
small and for each case we should investigate this, noting that it is not 
a quite straightforward concept.  

 We shall discuss $\psi$ as having developed into a 
classical stochastic field, Eq.(\ref{eq33}). The indeterminancy  represented 
 by the variables $e$ forces us to consider the average over the product 
of these variables so that we evaluate $<(\psi(x,\tau)^2>$ the result being
 \begin{equation}\label{VC1.1}
 \langle\psi(x,\tau)^2\rangle=
(2\pi)^{-3}\int d^3k \psi_k(\tau)\psi_k(\tau)^*,
\end{equation}
the same for every value of $x$. We require that $\sqrt{<(\psi(x,\tau)^2>}$    
be small compared with unity, since the relevant metric coefficient, 
Eq.\ref{eq24}, is $a(\tau)^2(1+2\psi)$. Our viewpoint is that in any 
particular  case satisfaction of this requirement 
forms sufficient justification for the perturbative
 approach, because the only way we can mount a comparison of the revised 
 metric coefficient with $a(\tau)^2$ is when we consider the basic equations 
 to be those in configuration space, and then indeed the revision is just a 
 perturbation through all space-time. It is true that if, having written the
 equations of motion in configuration space, we then project into $k$-space 
 to solve them, as we do, then certain of the $\psi(k)$ will be very large.
 But at this stage we are operating translation into $k$-space as a device 
 to aid solution. Also then the comparators from $a(\tau)$ will be much 
 larger, these translating either into a $k$-space $\delta$-function or, 
 in the non-ideal case, into a function very strongly peaked round the origin
 in $k$-space. 

 In the cases illustrated we see from Figure 3 that the requirement 
 is very well satisfied throughout and from Figure 4 that the 
 requirement is also satisfied though significantly better in the first
 part of reheating.

\subsection{Production of $\chi$ and Inflaton Particles}\label{VB}

It is a generic feature of this model, and probably of most other
similar models, that the inflaton particles or inhomogeneous field,
 $\varphi_1$, that enter the reheating period from inflation 
 attain an energy density comparable or greater than that of 
 the classical inflaton field, $\varphi_0$,during the preheating
 period and thereafter.
 In their equation of motion, (\ref{eq32}), we note the
 bilinear couplings to $\varphi_0$
 and the metric perturbation, $\psi$. Thus in preheating $\varphi_1$ 
 cannot be treated as a perturbation and we cannot neglect factors 
 like $\langle \varphi_1'^2 \rangle$.

 The occurence of parametric resonance for $\chi$ is as expected 
 sensitive to the value of $g$, and it is also, naturally, sensitive  
 to the value of $\Gamma_{\chi}$ which tends to inhibit it. In the
 case which we illustrate we see from Figure 1 that parametric 
 resonance occurs markedly but rather briefly and that afterwards the
 $\chi$ particles almost vanish under the influence of the friction 
 decay into a radiation fluid. As previously noted $\chi$ and $\psi$ have 
 independent stochastic variables.

\subsection{Magnitude of $\zeta$ During Reheating}\label{VC}

It has been found useful to define the parameter $\zeta$ of Eq.\ref{eq37} 
 where $\zeta_k \approx -{\cal R}_k$, the curvature perturbation \cite{LYTH,MUK}.
 \footnote{$\zeta$ being a linear function of the curvature perturbation
 $\psi$ arises with the same quantum operators as the non-homogeneous 
 inflaton field $\varphi_1$ as was noted for $\psi$ in section III. In  
 the successor classical stochastic variables the mode functions of
 $\varphi_1$, $\psi$ and $\zeta$ are complex but as we have noted
 previously can be converted to real by a time independent phase
 transformation \cite{PS}. It is this real quantity  which we quote here 
 for $\zeta$.}
For adiabatic perturbations with $k^2 \ll a^2H^2$, $\zeta_k$ is constant. 
 For example, in eras where there is only one scalar field or only
 one hydrodynamic fluid, then the perturbations are adiabatic. This holds 
 before the reheating period because 
  the scalar field $\chi$ is negligable, as we have seen above.
 However in the reheating period  we have two scalar fields 
 $\varphi_1,\chi$ and also the radiation fluid 
 (which arises from dissipative processes) 
 giving rise to an entropy perturbation, \cite{MOLL},and thus the
 theorem does not apply. Importantly, in addition, $\varphi_1$ can no 
 longer be treated as a perturbation and we do not do so as has been 
 emphasized in section \ref{VB}.

 In the equations of motion the wave number appears as $k^2$ and we 
 evaluate $\zeta$ for a value of $k$ such that $k^2 < 10^{-4}a^2H^2$
 throughout the reheating period. We see from Figs.3 and 5 that 
$\zeta$ has in the radiation era
  modestly decreased from its value at the end of the inflationary era.
  When we switch off the dissipative decay into a fluid, so that there is
  preheating without reheating, $\zeta$ now increases markedly as
 shown in Figs. 4 and 6. We also see from Figs. 5 and 6 that $\zeta$ 
 is initially constant as it should be since $\varphi_1$ 
 is then small enough to be treated to first order only.

 \section{Summary and Conclusions}\label{VI}

 We have studied a single field ($\varphi$) model of inflation, with
 another scalar field ($\chi$) which is naturally quiescent during 
 inflation and becomes active post inflation with the possibility 
 of parametric resonance and preheating as the classical part,
 $\varphi_0$, of the inflaton field decreases and oscillates. We have
 included reheating through friction-type decay mechanisms of the
 scalar  fields into a gas of unspecified relativistic particles. We have
 adopted a method having stochastic variables naturally succeeding the 
 quantum operators of the early inflation era and have developed 
 equations for the post-inflation period by taking averages over the 
 stochastic bilinear forms in the scalar fields. This enables a 
 non-perturbative treatment of the stochastic scalar fields 
 $\chi$ and $\varphi_1$ (where $\varphi = \varphi_0 +\varphi_1$) 
 though the metric perturbation field, $\psi$, has to be treated 
 perturbatively.

 We find that the non-homogeneous scalar field (which might also
 be described as bearing inflaton particles), $\varphi_1$, remains 
 strong with an energy density which competes (according to the  
 values of the decay parameters) with that of the gas of relativistic 
 particles, during a large part of the reheating period.

 There are cases of the parameters in which the metric perturbation 
 $\psi$ markedly resonates, at much the same time as a parametric
 resonance in  $\chi$, and exceeds the perturbative limit. We 
 have to reject such cases from consideration though we cannot say 
 whether the failure be mathematical or physical. There are, however, 
 many valid cases.

 The curvature parameter, $\zeta$, for wavelengths relevant to the 
 cosmic microwave background fluctuations occurs very conveniently 
 as a constant in  non-entropic single field inflation models.
 In this present more complicated model it changes from its value 
 in the inflationary period by up to a factor of the order of 10 in 
 the examples presented in this paper.

 We thank Andrew Liddle, David Lyth and Luis Mendes for valuable 
 conversations.

\begin{figure}
\begin{center}
\input{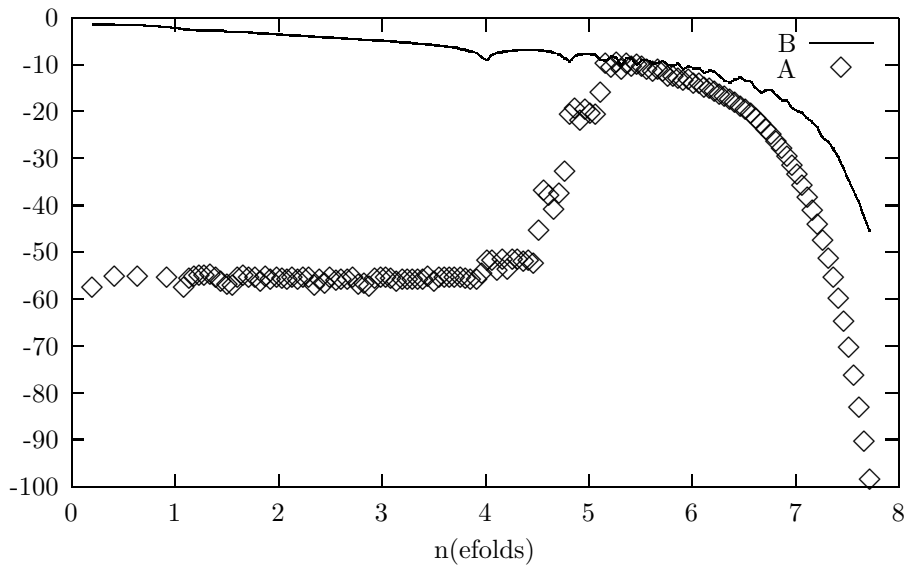}
\end{center}
\caption{Logarithmic plot of energy densities of $\chi$ (A)and
 $\varphi_0$ (B) versus the number of e-folds of expansion 
 in the reheat era. The parameters are $g/m=10^4,\Gamma_{\chi}/m=0.01,
 \Gamma_{\varphi}/m=0.004$, with other constants as given in 
 the text, Section \ref{V}. In all figures densities are in units
 $m^2m_{Planck}^2$, where $m=10^{-7}m_{Planck}$.}   
\end{figure}
 
\begin{figure}
\begin{center}
\input{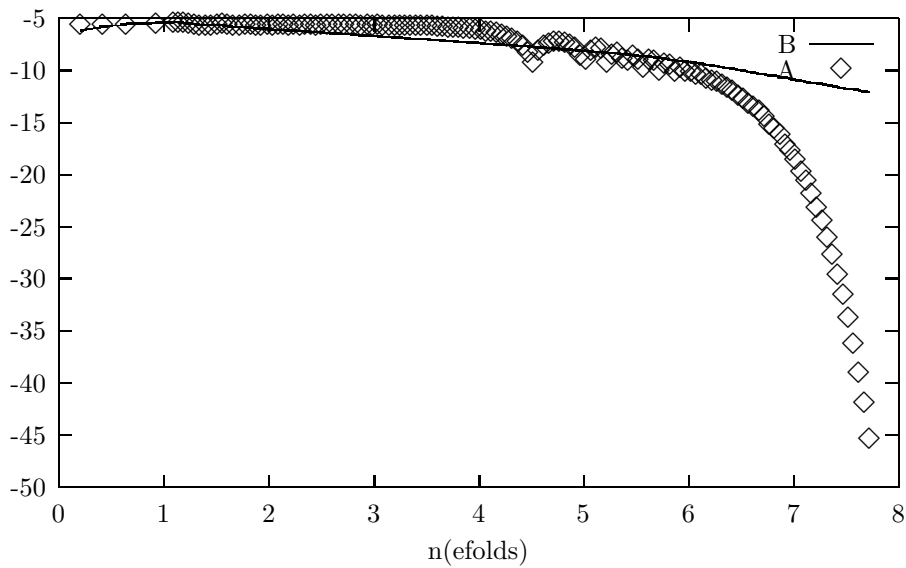}
\end{center}
\caption{$\log \rho =A$ and  $\log({\rm energy density}(\varphi_1)) = B$
 versus the number of e-folds of expansion in reheating.$\rho$ is the
 density of the radiation gas and $\varphi_1$ is the inhomogeneous 
 (or particle) part of the inflaton field. The parameters are as for 
 Figure 1.}
 \end{figure}

\begin{figure}
\begin{center}
\input{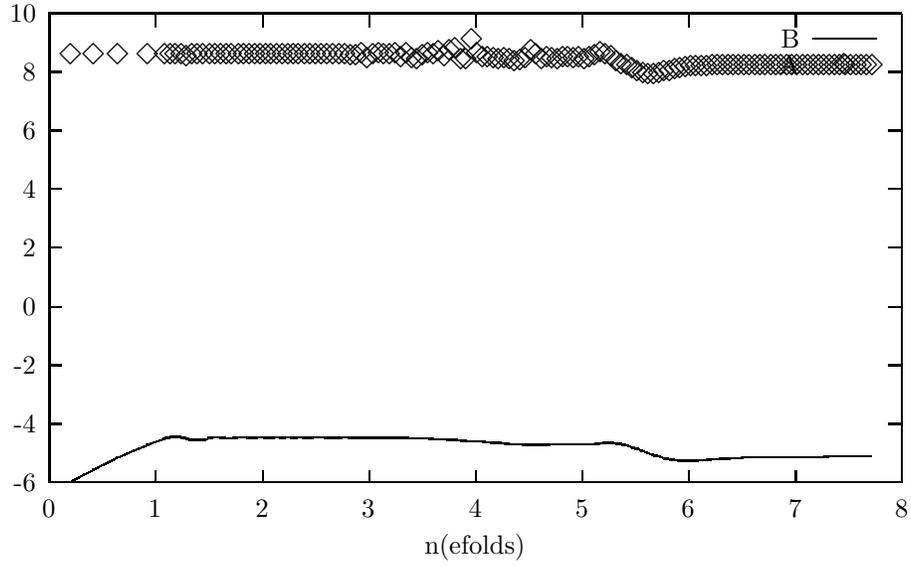}
\end{center}
\caption{$\log(\langle |\psi^2|\rangle) = A$ and $\log(\zeta) = B$ 
 where $\psi$ is the metric perturbation and $\zeta$ is defined 
 in the text. The parameters are as for  Figure 1.}
 \end{figure}

\begin{figure}
\begin{center}
\input{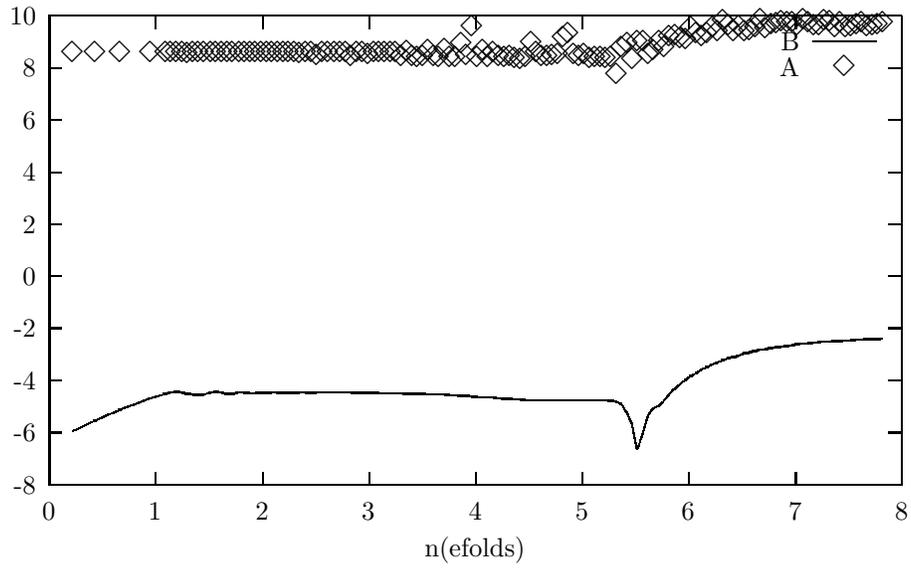}
\end{center}
\caption{As for Figure 3, but with $\Gamma_{\chi}=0, \Gamma_{\varphi}=0$ }
 \end{figure}

\begin{figure}
\begin{center}
\input{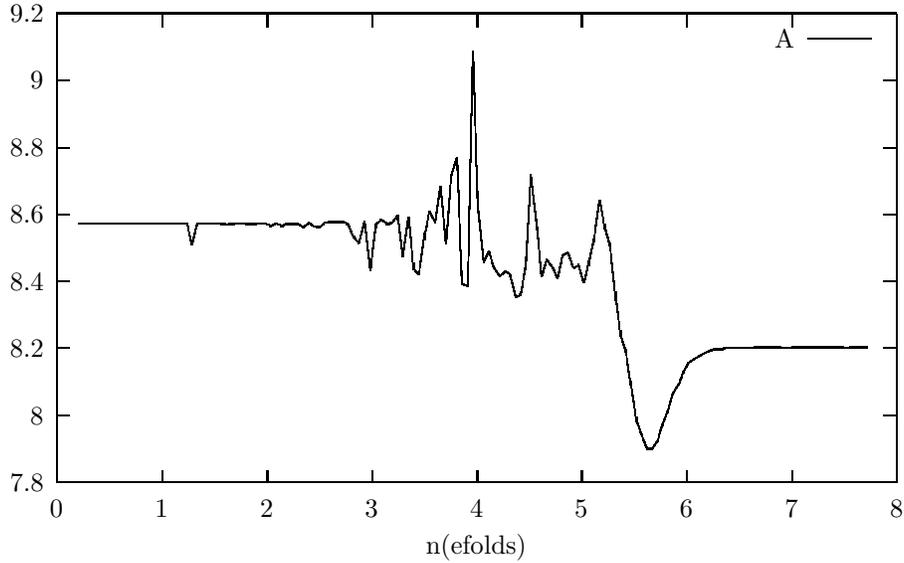}
\end{center}
\caption{$\log(\zeta)$ of  Figure 3 shown in more detail for the first 
 8 efolds of reheating.}
 \end{figure}

\begin{figure}
\begin{center}
\input{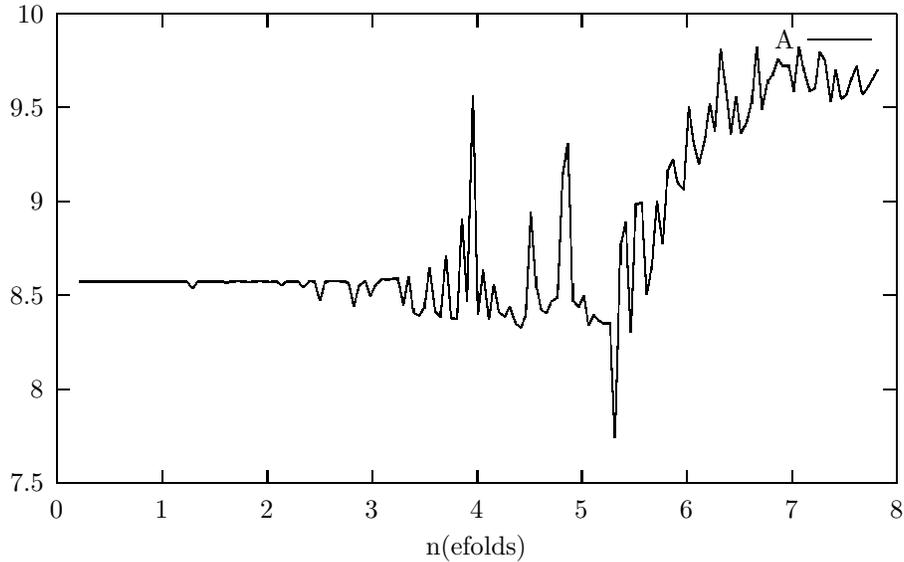}
\end{center}
\caption{$\log(\zeta)$ of  Figure 4 shown in more detail for the first 
 7 efolds of reheating. Comparison of this with Figure 5 illustrates
 differences in $\zeta$ in the cases with and without
 frictional decay of the scalar fields.}
 \end{figure}

\appendix
\section{Bianchi Identity and the Frictional Decay Terms.}
Working to first order in the metric perturbation $\psi$ a  Bianchi
identity gives
 \begin{equation}\label{A1}
D_{\nu} T^{\nu}_0 \equiv E_{\rho}+\varphi'S_{\varphi}+\chi'S_{\chi}=0
 \end{equation} 
 \begin{equation}\label{A2}
 E_{\rho} \equiv \rho'+(\rho+p)(3\alpha-3\psi'+
 (\partial_i-2\psi_{,i})u_0\partial u^i)            
 \end{equation} 
 \begin{equation}\label{A3}
 S_{\varphi}=(1-2\psi)(\varphi''+3\alpha\varphi'-3\psi'\varphi')-
 \psi'\varphi'+a^2\bar m ^2\varphi-(1+2\psi)\varphi_{,i,i}            
 \end{equation} 
\begin{equation}\label{A4}
 S_{\chi}=(1-2\psi)(\chi''+3\alpha\chi'-3\psi'\chi')-
 \psi'\chi'+a^2\bar M ^2\chi-(1+2\psi)\chi_{,i,i}            
 \end{equation} 
 where, from the space-time Einstein equation, the fluid velocity, $u^i$,
 is given by
 \begin{equation}\label{A5}
 a^{-2}(\rho+p)u_0 u^i = -\varphi'\varphi_{,i}+
 (4\pi G)^{-1}(\psi'+\alpha\psi).            
 \end{equation} 
 Equating the expressions in (\ref{A2},\ref{A3},\ref{A4}) to zero 
 separately gives the fluid equation and the scalar equations. To add 
 friction we use Eq.(\ref{A1}) to write
  \begin{equation}\label{A6}
\lbrack E_{\rho} -\Gamma_{\varphi}(\varphi')^2-\Gamma_{\chi}(\chi')^2\rbrack
+\varphi'\lbrack S_{\varphi}+\Gamma_{\varphi}\varphi' \rbrack
+\chi'\lbrack S_{\chi}+\Gamma_{\chi}\chi' \rbrack =0.
 \end{equation} 
  This equation, which is equivalent to the Bianchi identity, is solved 
 by equating each term in square brackets to zero. This modifies the 
 fluid and scalar equations to give a transfer of fluid and scalar densities
 which however is modulated by the complexity of our complete set of 
 equations. In our model $\rho=\rho_0+\rho_1,\varphi=\varphi_0 =
 \varphi_1$ where $\rho_0,\varphi_0$ are homogeneous and 
 $\rho_1,\varphi_1$ are non-homogeneous and along with $\psi$ linear in
 the same stochastic variables - this being forced through consistency
 of the various equations of the theory; $\chi$ is linear in different
 stochastic variables. Each of the first two equations can be divided into 
 a homogeneous part and a part linear in the stochastic variables with the 
 occurence of quadratic or cubic stochastic variable terms being
 dealt with by    
 appropriate ensemble averaging. Thus each of the first two equations 
 is divided into two two halves which must separately be zero and 
 after some cross substitution we get, together with the equation 
 from the third square bracket the equations for 
 $\rho_0,\rho_1, \varphi_0,\varphi_1,$ and $\chi$ which appear in the text.

\section{Evaluation of the Loop Integrals.}
Integrals such as that in Eq.(\ref{eq19}) occur throughout the equations of 
motion and are evaluated numerically and we have to adopt
 a finite range of wave number, $k$. If the integrals diverge as 
$k \rightarrow \infty$ then the upper limit of the $k$ integration forms a
cut-off which is the crudest way of dealing with such ultra-violet 
divergences. Two points of view may be taken on this. Firstly this may be 
considered equivalent to a renormalization procedure in mass and other 
quantities. Secondly the Lagrangian used may be considered as an effective 
Lagrangian which has absorbed extra degrees of freedom coming from 
supersymmetry which eliminate divergences at higher momentum, the cut-off 
representing this effect. We have taken a cut-off to correspond to a 
wavelength of $H^{-1}$ so that $k_{cutoff} \approx 2\pi aH$. 

Also an infra-red cutoff is needed when the spectral index $n$ is less than 
 1 irrespective of the particular inflation model (of the normal type); for
 the power law inflation model $n$ is always less than 1. Inflationary 
 theories which are in agreement with COBE observations have a value of
 $n$ near, either greater or less than, 1 and thus there is a steep rise 
 of (for a principal example) $\varphi_k$ as $k\to 0$ as mentioned above in 
 section \ref{IVB4}. First we should make the general remark, independent 
 of any particular value of $n$, that for $k=0$ the inflation era 
 perturbation is homogeneous and may be regarded as incorporable in the 
 classical homogeneous field $\varphi_0$. The problem arises that for $k$ 
 {\it near} to $0$ the  perturbation is {\it nearly} homogeneous and also 
 a singularity of the loop  integrand can arise at the end of 
 inflation and in reheat. When the actual integral diverges the problem 
 is acute and we deal with it by an infra-red cut-off taking the lower
 limit of the integration as $k=k_0$ so that  Eq.\ref{eq17} becomes 
 $\langle\varphi_1^2\rangle=(2\pi)^{-3}\int_{k_0} d^3k \varphi_k\varphi_k^*$.
 The perturbations for $k<k_0$ we regard as hidden by renormalization of the 
 field $\varphi_0$. (In the equations of motion $\langle\varphi_1^2\rangle$
 always occurs in association with $\varphi_0^2$.) 
 $k_0$ should be small enough so that a $\varphi_k$ which
 gives rise to observable effects is explicitly included in the loop
 integral. Thus $k_0<k_m\equiv a_mH_m$ where the suffix $m$ denotes the value 
 at the time when the matter era begins,as observations of the CMBR 
 fluctuations are possible at $k$-values nearly of this order; $k_m$ is tiny 
 compared with the corresponding value of $aH$ at reheat. 

 Now consider 
 any value of $k$ such that $k \ll aH$ throughout the reheat era; then
 $\varphi_k \propto k^{n/2-2}$ at the end of inflation and since for such 
 small values of $k$ (wavelength much greater than the Hubble radius) the 
 $k^2$ term in the reheating equations of motion can be neglected means 
 that the initial proportionality is maintained throughout reheat and thus
 $\varphi_k = \tilde\varphi(\tau)k^{n/2-2}$. We thus evaluate the loop 
 integral as
  \begin{equation}\label{IVC.1}
\langle\varphi_1^2\rangle=
(2\pi^2)^{-1}\int_{k_0}^{k_1} \varphi(\tau)\varphi(\tau)^*k^{n-2}dk+
(2\pi)^{-3}\int_{k_1} d^3k \varphi_k\varphi_k^*
\end{equation}
 where $k_1 \ll aH$ throughout reheat and the last term can be evaluated
 approximately using discrete values of the integrand since the delicate 
 feature is included in the first term on the right hand side.

We shall now treat these small $k$ contributions in more detail, 
 distinguishing the two cases $n<1$ and $n>1$.

{\bf Case 1:}$n<1$. Let $n=1-\epsilon$ where $\epsilon>0$. The analytic 
integration in Eq.\ref{IVC.1} is
  \begin{equation}\label{IVC.2}
\int_{k_0}^{k_1} k^{-1-\epsilon}dk=
 (k_0^{-\epsilon}-k_1^{-\epsilon})/\epsilon.
\end{equation}

Because $k_0=O(k_m)$ and $\epsilon=O(.1)$ the first term on the right of 
 Eq.\ref{IVC.1} contributes a relatively large positive quantity to 
 the loop integral. For the power-law inflation model $n<1$ so this is 
 the case that applies and is used in this paper.

{\bf Case 2:} $n>1$. Let $n=1+\epsilon$ where $\epsilon>0$.
 \begin{equation}\label{IVC.3}
\int_{k_0}^{k_1} k^{-1+\epsilon}dk=
(k_1^{\epsilon}-k_0^{\epsilon})/\epsilon 
 \approx  k_1^{\epsilon}/\epsilon.
\end{equation}
 So the corresponding contribution is much smaller than that in the 
 preceeding case which applies in this paper.

\end{document}